\documentclass[10pt,twocolumn,letterpaper]{article}

\usepackage{iccv}
\usepackage{times}
\usepackage{epsfig}
\usepackage{graphicx}
\usepackage{amsmath}
\usepackage{amssymb}
\usepackage{subcaption}
\usepackage[table]{xcolor}

\usepackage{amsfonts}
\usepackage{amsxtra}
\usepackage{amsthm}
\usepackage{mathtools}
\usepackage{relsize}
\usepackage{array}
\usepackage{bm}

\usepackage{float}
\usepackage{enumitem}

\usepackage{multirow}
\usepackage{hhline}
\usepackage{siunitx}
\usepackage{fancyhdr}
\usepackage{fancyvrb}
\usepackage{algorithm}
\usepackage{algpseudocode}
\usepackage{longtable}
\usepackage{xspace}
\usepackage{tocloft}
\usepackage{sectsty}
\usepackage[pagebackref=true,breaklinks=true,letterpaper=true,colorlinks,bookmarks=false]{hyperref}
\usepackage[space]{grffile}
\usepackage{bookmark}
 \iccvfinalcopy % *** Uncomment this line for the final submission

\def\iccvPaperID{9} % *** Enter the ICCV Paper ID here

\ificcvfinal\fi

\begin{document}

%%%%%%%%% TITLE
\title{\ Edge-Informed Single Image Super-Resolution}

\author{Kamyar Nazeri, Harrish Thasarathan, ~and~ Mehran Ebrahimi\\
University of Ontario Institute of Technology, Canada\\
{\tt\small {kamyar.nazeri@uoit.ca~~~ harrish.thasarathan@uoit.net  ~~~  mehran.ebrahimi}@uoit.ca}\\
\small{\url{http://www.ImagingLab.ca}}
}

\maketitle
\ificcvfinal\thispagestyle{empty}\fi

%%%%%%%%% ABSTRACT
\begin{abstract}

The recent increase in the extensive use of digital imaging technologies has brought with it a simultaneous demand for higher-resolution images. We develop a novel ``edge-informed'' approach to single image super-resolution (SISR).  The SISR problem is reformulated as an image inpainting task. We use a two-stage inpainting model as a baseline for super-resolution and show its effectiveness for different scale factors ($\times 2$, $\times 4$, $\times 8$) compared to basic interpolation schemes. This model is trained using a joint optimization of image contents (texture and color) and structures (edges). Quantitative and qualitative comparisons are included and the proposed model is compared with current state-of-the-art techniques. We show that our method of decoupling structure and texture reconstruction improves the quality of the final reconstructed high-resolution image.
\end{abstract}

%%%%%%%%% BODY TEXT
\section{Introduction}

Super-Resolution (SR) is the task of inferring a high-resolution (HR) image from one or more given low-resolution (LR) images. SR plays an important role in various image processing tasks with direct applications in medical imaging, face recognition, satellite imaging, and surveillance \cite{farsiu2004advances}. Many existing SR methods reconstruct the HR image by fusing multiple instances of a LR image with different perspectives. These are called Multi-Frame Super-Resolution methods \cite{farsiu2004fast}. However, in most applications, only a single instance of the LR image is available from which missing HR information needs to be recovered.  Single-Image Super-Resolution (SISR) is a challenging ill-posed inverse problem \cite{ebrahimi2007solving} that normally requires prior information  to restrict the solution space of the problem \cite{shi2016real}.

We take inspiration from a recent image inpainting technique introduced by Nazeri \etal \cite{nazeri2019edgeconnect} to propose a novel approach to Single-Image Super-Resolution by reformulating the problem as an in-between pixels inpainting task. Increasing the resolution of a given LR image requires recovery of pixel intensities in between every two adjacent pixels. The missing pixel intensities can be considered as missing regions of an image inpainting problem. Our inpainting task is modelled as a two stage process that separates structural inpainting and textural inpainting to ensure high frequency information is preserved in the recovered HR image. The pipeline involves first creating a mask for every extra row and column that needs to be filled in the reconstruction of the HR image. The edge generation stage then focuses on ``hallucinating'' edges in missing regions, and the image completion stage uses the hallucinated edges as prior information to estimate pixel intensities in the missing regions. 

\begin{figure}[h]
	\centering
	\begin{subfigure}{.14\textwidth}
		\centering
		\includegraphics[width=\textwidth]{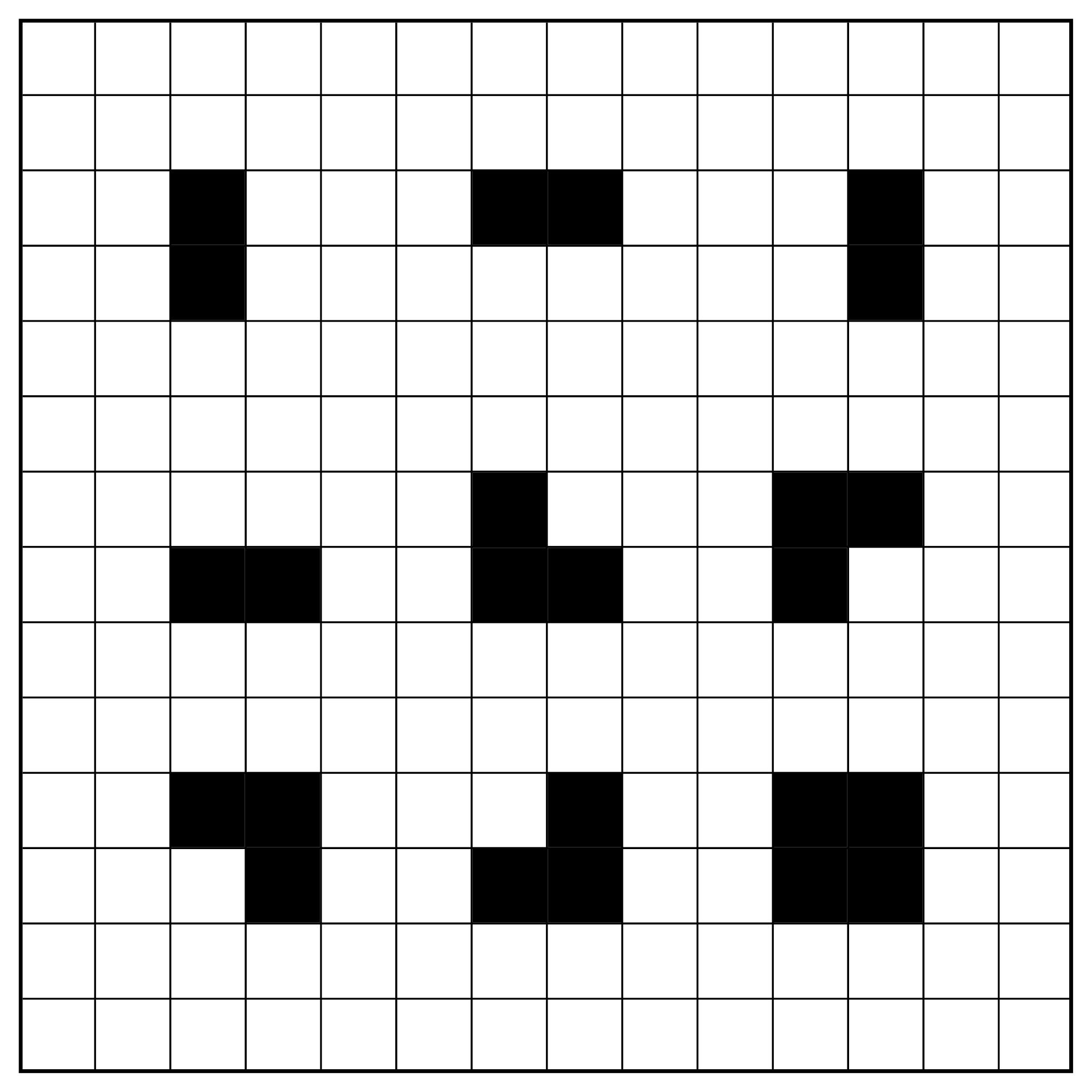}
		\caption{Ground Truth}
	\end{subfigure}
	\begin{subfigure}{.14\textwidth}
		\centering
		\includegraphics[width=\textwidth]{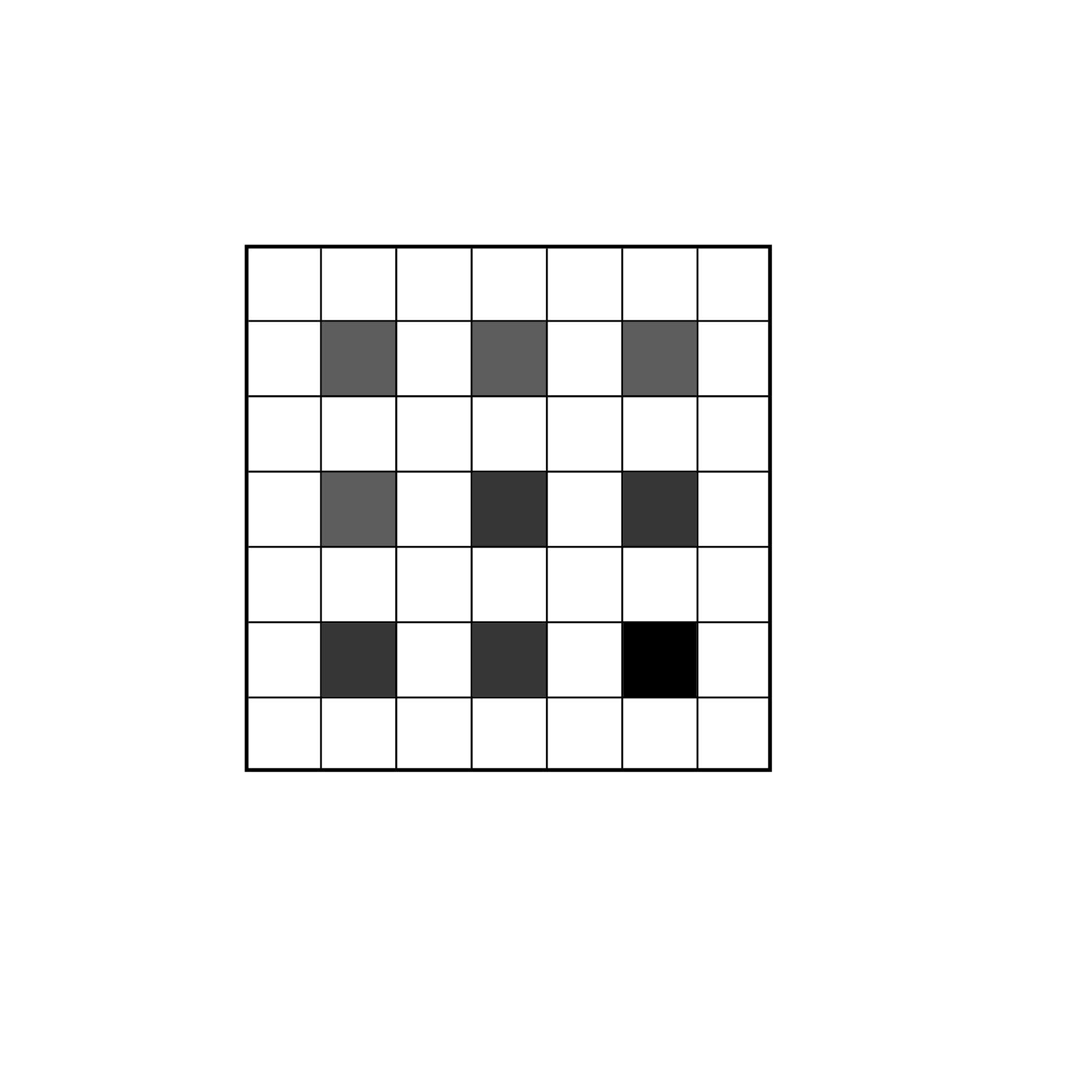}
		\caption{LR Image}
	\end{subfigure}
	\begin{subfigure}{.14\textwidth}
		\centering
		\includegraphics[width=\textwidth]{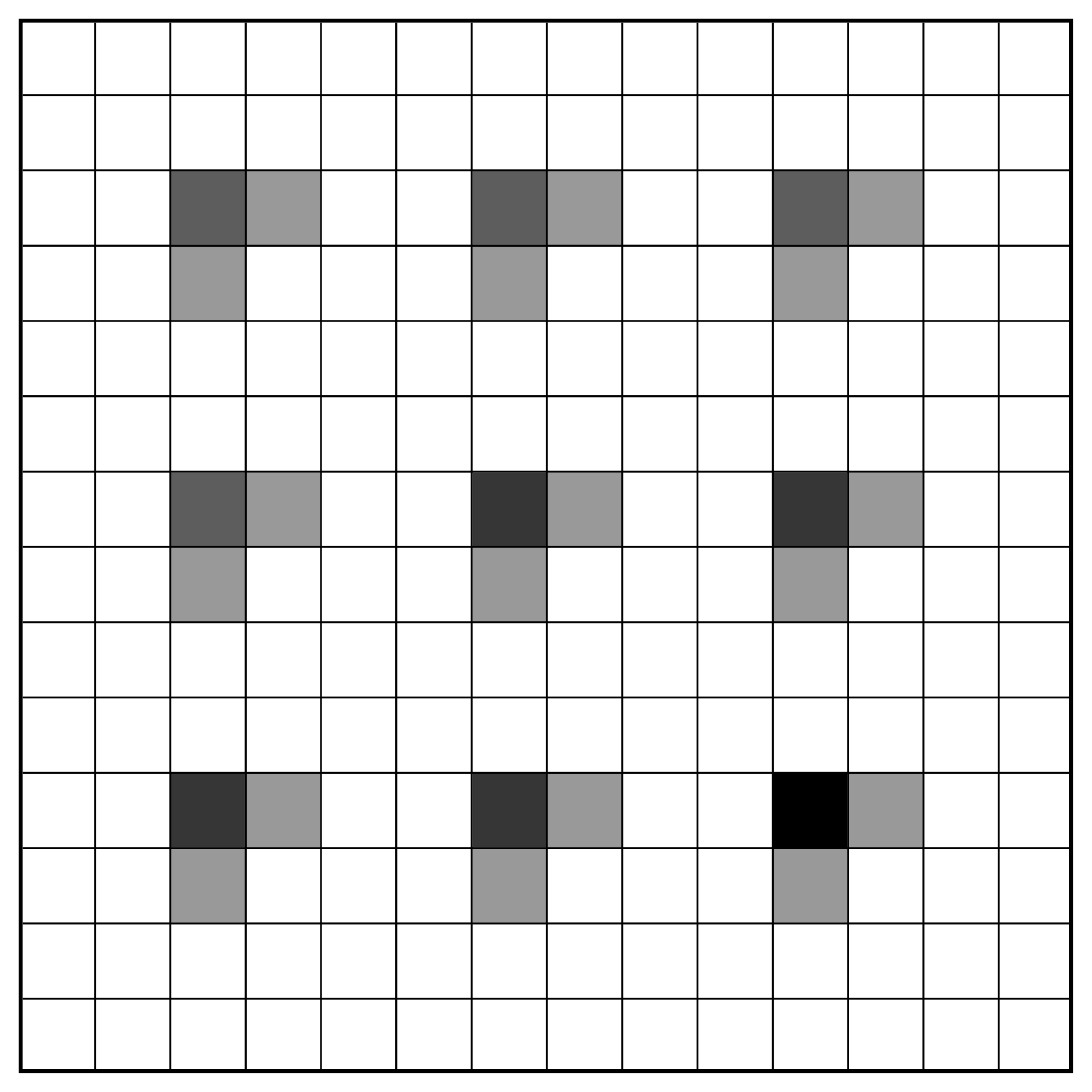}
		\caption{HR Estimate}
	\end{subfigure}
	\caption{Schematic illustration of the super-resolution problem. (a) The ground truth image, (b) The image downsampled by a factor of two. Each four-pixel segment of information on the left turn into one pixel in the middle, as a result, the structure and orientation of edges are not distinguished anymore as the problem is ill-posed. (c) The reconstruction of a high-resolution image from one-pixel segments of information using bilinear interpolation. Most distinctive features in the original image are lost and the result is blurry around the edges. }
	\label{fig:sr}
\end{figure}

\begin{figure}[h]
	\centering
	\begin{subfigure}{.14\textwidth}
		\centering
		\includegraphics[width=\textwidth]{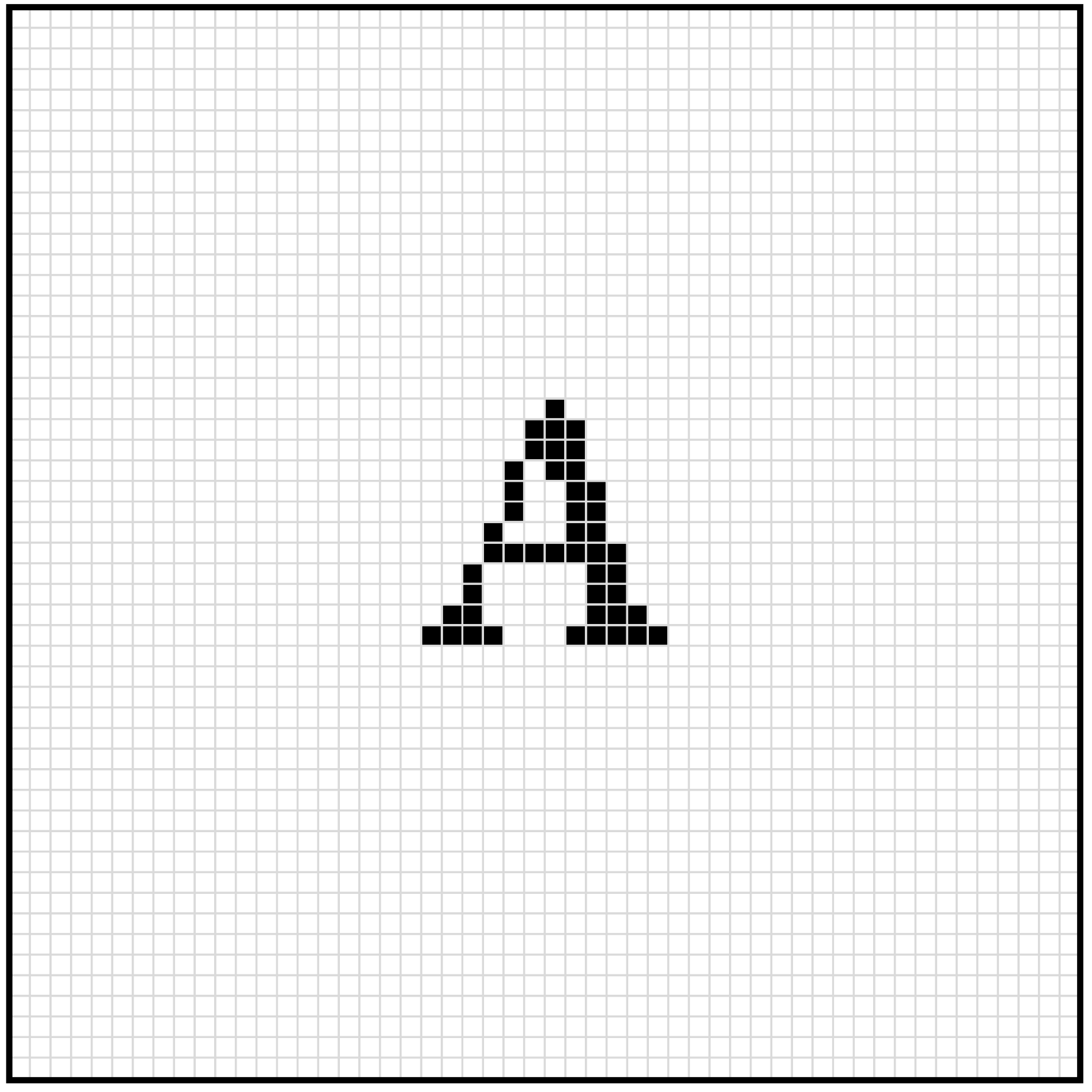}
		\caption{LR image}
	\end{subfigure}
	\begin{subfigure}{.14\textwidth}
		\centering
		\includegraphics[width=\textwidth]{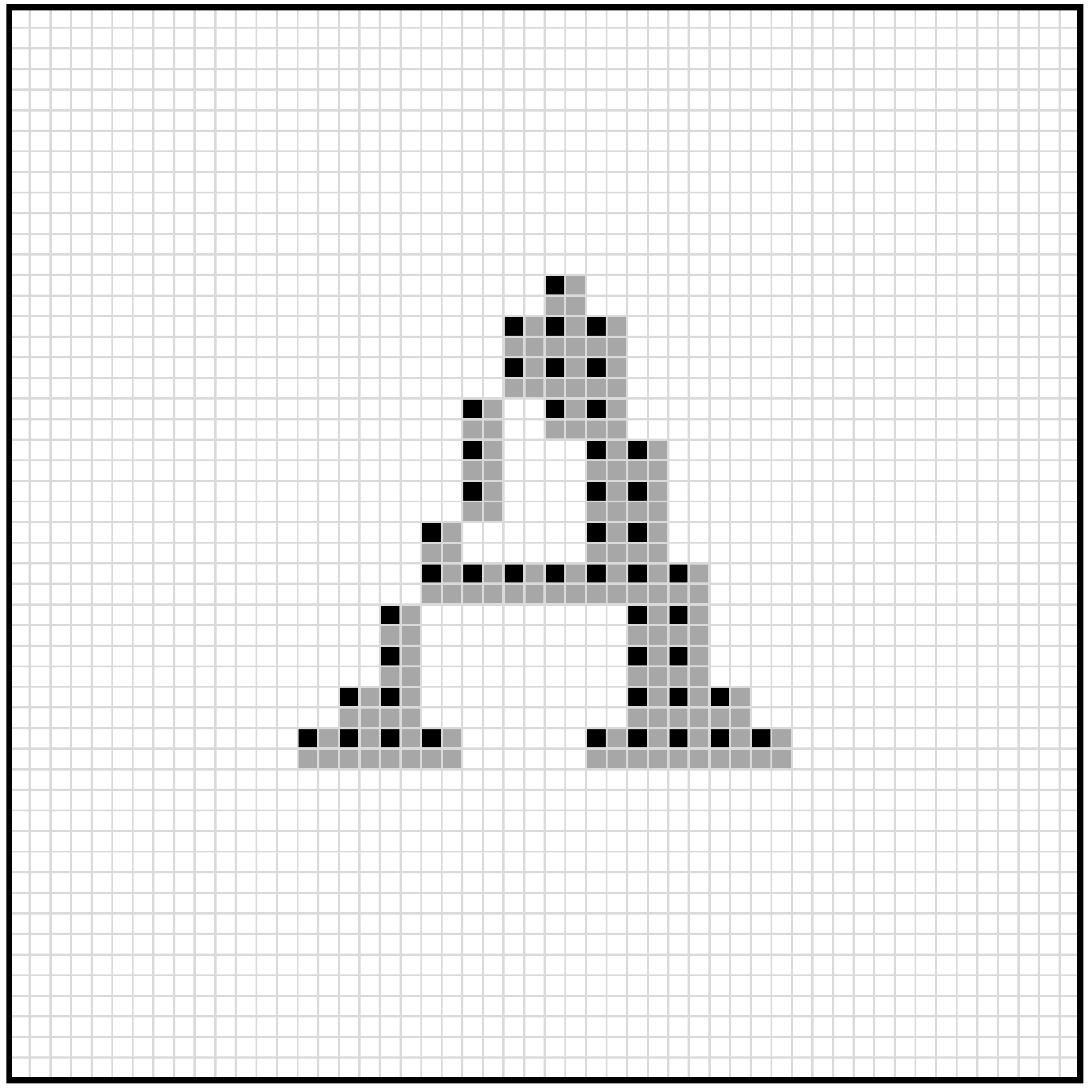}
		\caption{Upsample $2 \times$}
	\end{subfigure}
	\begin{subfigure}{.14\textwidth}
		\centering
		\includegraphics[width=\textwidth]{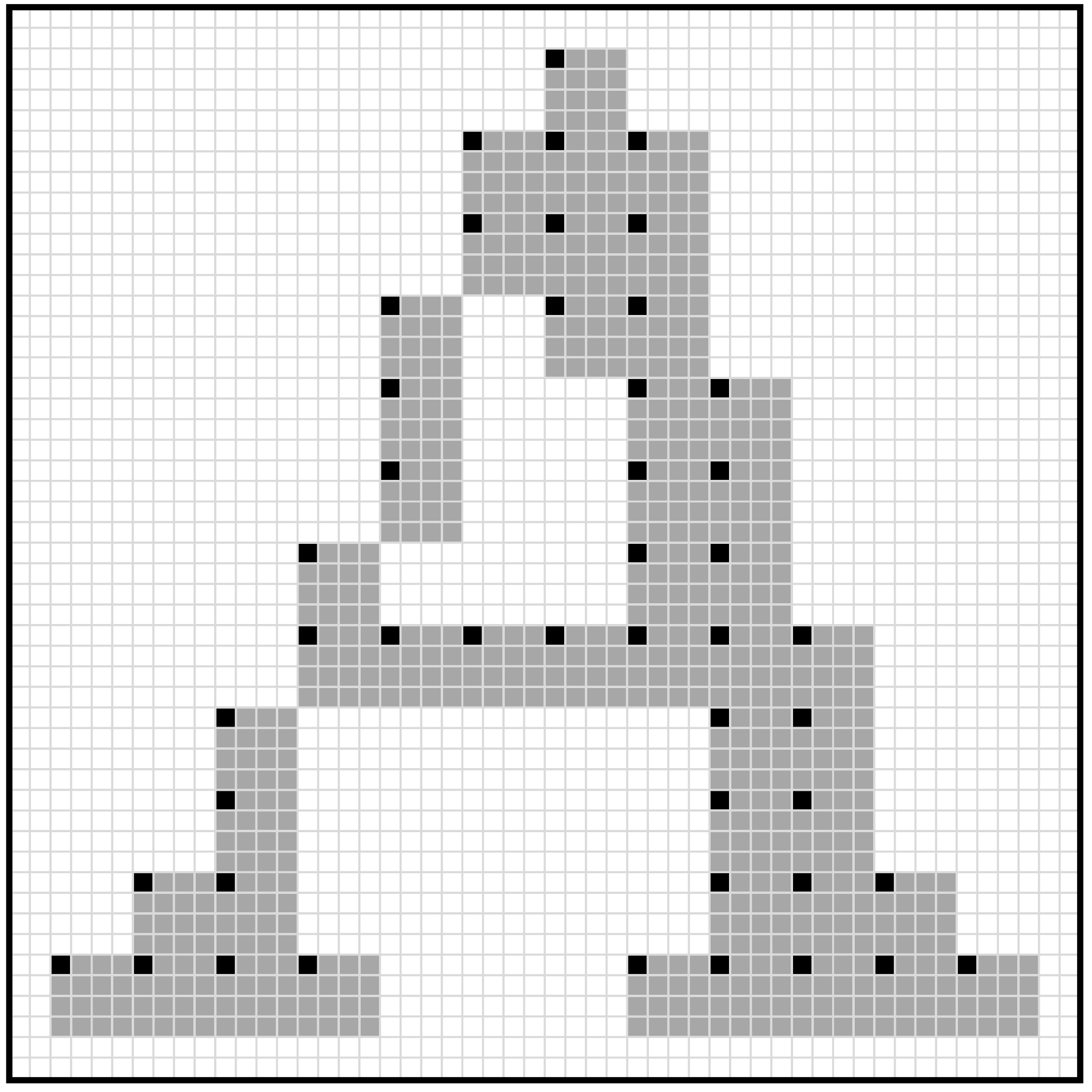}
		\caption{Upsample $4 \times$}
	\end{subfigure}
	\caption{An illustration of the proposed inpainting-based method for SISR. (a) The original LR image. (b) Upsampling by a factor of two corresponds to interpolating one pixel between every two adjacent pixels. We add an extra empty row and column for every row and column in the ground truth image (shown in gray) which we fill by an inpainting process. (c) Upsampling by a factor of four corresponds to interpolating three pixels between every two adjacent pixels where we can add three extra empty rows and columns for every row and column in the ground truth image to be inpainted.}
	\label{fig:sr_inpainting}
\end{figure}

\section{Related Work}
Many approaches to SISR have been presented in literature. These methods have been extensively organized by type according to their image priors in a study by Yang \etal \cite{yang2014single}. \textbf{Prediction models} generate HR images through predefined mathematical functions. Examples include bilinear interpolation and bicubic interpolation \cite{de1978practical}, and Lanczos resampling \cite{duchon1979lanczos}. \textbf{Edge-based methods} learn priors from features such as width of an edge \cite{fattal2007image}, or parameter of a gradient profile \cite{sun2008image} to reconstruct the HR image. \textbf{Statistical methods} exploit different image properties such as gradient distribution \cite{shan2008fast} to predict HR images. Patch-based methods use exemplar patches from external datasets \cite{chang2004super,freeman2002example} or the image itself \cite{irani2009super,freedman2011image} to learn mapping functions from LR to HR.

\textbf{Deep Learning-based methods} have achieved great performance on SISR using deep convolutional neural networks (CNN) with a per-pixel Euclidean loss \cite{shi2016real,dong2014learning,kim2016accurate}. 
%,kim2016deeply
Euclidean loss, however, is less effective to reconstruct high-frequency structures such as edges and textures. Recently Johnson \etal \cite{johnson2016perceptual} proposed feed-forward CNN using a perceptual loss. In particular, they used a pre-trained VGG network \cite{simonyan2014very} to extract high-level features from an image effectively separating content and style. Their model was trained with a joint optimization of \textit{Feature reconstruction loss} and \textit{Style reconstruction loss} and achieved state-of-the-art results on SISR for challenging $\times 8$ magnification factor. To encourage spatial smoothness and mitigate the checkerboard artifact \cite{odena2016deconvolution} of using feature reconstruction loss, they introduced \textit{total variation regularization} \cite{rudin1992nonlinear} to their model objective. Sajjadi \etal \cite{sajjadi2017enhancenet} proposed to use style loss in a patch-wise fashion to reduce the checkerboard artifact and enforce locally similar textures between the HR and ground truth images. They also used an adversarial loss to produce sharp results and further improve SISR results. Adversarial loss has also shown to be very effective in producing realistically synthesized high-frequency textures for SISR \cite{ledig2017photo,haris2018deep,park2018srfeat}, however, the results of these GAN-based approaches tend to include less meaningful high-frequency noise around the edges that is unrelated to the input image \cite{park2018srfeat}. Our work herein is inspired by the model proposed by Liu \etal \cite{Liu_2018_ECCV} which extended their image inpainting framework to image super-resolution tasks by offsetting pixels and inserting holes. We present a SISR model that simultaneously improves structure, texture, and color to generate a photo-realistic high-resolution image.

\section{Model}
We propose a Single Image Super-Resolution framework based on a two stage adversarial model \cite{goodfellow2014generative} consisting of an edge enhancement step and an image completion step. Both the edge enhancement and image completion steps consist of their own generator/discriminator pair that decouples SISR into two separate problems \textit{i.e.} structure and texture. Let $G_1$ and $D_1$ be the generator and discriminator for the edge enhancement step, and $G_2$ and $D_2$ be the generator and discriminator for the image completion step. Our edge enhancement and image completion generators are built from encoders that downsample twice, followed by eight residual blocks \cite{he2016deep}, and decoders that upsample to the original input size. We use dilated convolutions in our residual layers. 
%which creates a receptive field of size $205$ at the final residual block. 
Our generators follow similar architectures to the method proposed by Johnson \etal \cite{johnson2016perceptual} shown to achieve superior results for super-resolution \cite{sajjadi2017enhancenet, gondal2018unreasonable}, image-to-image translation \cite{zhu2017unpaired}, and style transfer. Our discriminator follows the architecture of a $70 \times 70$ PatchGAN \cite{isola2017image, zhu2017unpaired} that classifies overlapping $70 \times 70$ image patches as real or fake. We use instance normalization \cite{ulyanov2017improved} across all layers of the network, which normalizes across the spatial dimension to generate qualitatively superior images during training and at test time. 

\begin{figure*}
	\centering
	\includegraphics[height=.14\textheight]{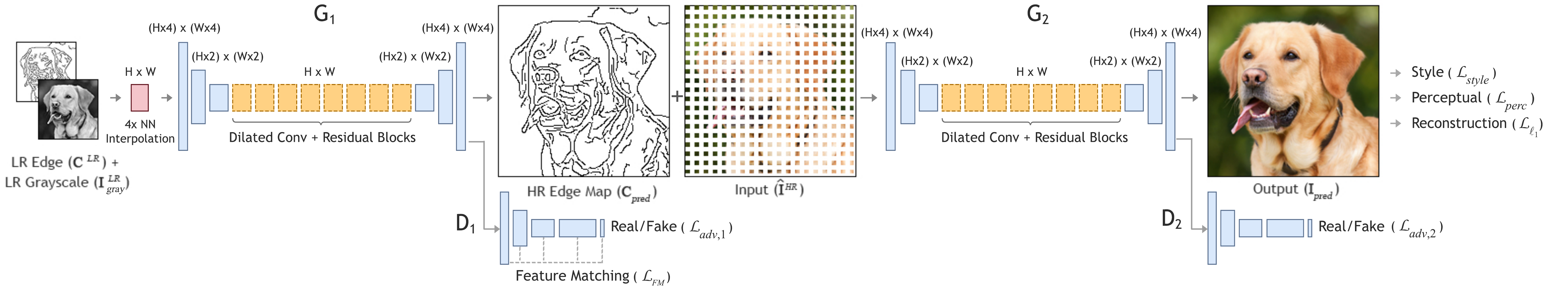}
	\caption{Summary of our proposed method. $G_1$ takes a low resolution greyscale image $\mathbf{I}^{LR}_{gray}$ and its corresponding low resolution edge map $\mathbf{C}^{LR}$ interpolated to the desired high resolution image size and outputs a high resolution edge map $\mathbf{C}_{pred}$. $G_2$ takes the high resolution edge map generated by $G_1$ as well as an incomplete HR image $\mathbf{I}_{gt}$ created by offsetting the pixels of the ground truth LR image using a fixed fractionally strided convolution kernel. The output is the high resolution image $\mathbf{I}_{pred}$.}
	\label{seq3}
\end{figure*}

\subsection{Edge Enhancement}
Our edge enhancement stage boosts the edges obtained from a low-resolution image to yield a high-resolution edge map. Let $\mathbf{I}^{LR}$ and $\mathbf{I}^{HR}$ be the low-resolution and high-resolution images. Their corresponding edge maps will be denoted as $\mathbf{C}^{LR}$ and $\mathbf{C}^{HR}$ respectively and $\mathbf{I}^{LR}_{gray}$ is a grayscale counterpart of the low-resolution image. We add a nearest-neighbor interpolation module at the beginning of the network to resize the low-resolution image and its Canny edge-map to the same size as the HR image. The edge enhancement network $G_{1}$ predicts the high-resolution edge map
\begin{equation}
    \mathbf{C}_{pred} = G_1(\mathbf{I}^{LR}_{gray},\mathbf{C}^{LR}),
\label{eq:HR_edgemap}
\end{equation}
where $\mathbf{I}^{LR}_{gray}$ and $\mathbf{C}^{LR}$ are the inputs to the network. The hinge variant \cite{miyato2018spectral} of the adversarial loss objective over the generator and discriminator are defined as
\begin{equation}
	\mathcal{L}_{G_1} = - \mathbb{E}_{\mathbf{I}_{gray}} \left[ D_1 (\mathbf{C}_{pred}, \mathbf{I}_{gray}) \right],
	\label{eq:g1loss}
\end{equation}
\begin{multline}
	\mathcal{L}_{D_1} = \mathbb{E}_{(\mathbf{C}_{gt},\mathbf{I}_{gray})} \left[ \max (0, 1 - D_1 (\mathbf{C}_{gt}, \mathbf{I}_{gray})) \right] \\
	+ \mathbb{E}_{\mathbf{I}_{gray}} \left[ \max (0, 1 + D_1 (\mathbf{C}_{pred}, \mathbf{I}_{gray})) \right].
	\label{eq:d1loss}
\end{multline}
We also include a feature matching loss objective $\mathcal{L}_{FM}$ \cite{wang2018high} to our edge enhancement generator which compares activation maps in the intermediate layers of the discriminator. This stabilizes the training process by forcing the generator to produce results with representations that are similar to real images. Perceptual loss \cite{johnson2016perceptual,gatys2016image,gatys2015texture} has also been known to accomplish this same task using a pretrained VGG network. However, since the VGG network is not trained to produce edge information, it fails to capture the result that we seek in the initial stage. The feature matching loss is defined as

\begin{multline}
	\mathcal{L}_{FM} = \mathbb{E}\left[ \sum_{i} \frac{1}{N_i} \left\lVert D^{(i)}_1 (\mathbf{C}_{gt})- D^{(i)}_1 (\mathbf{C}_{pred}) \right\rVert_1 \right],
\end{multline}
where $N_i$ is the number of elements in the $i$'th activation layer, and $D_1^{(i)}$ is the activation in the $i$'th layer of the discriminator. Spectral normalization (SN) \cite{miyato2018spectral} further stabilizes training by scaling down weight matrices by their respective largest singular values, effectively restricting the Lipschitz constant of the network to one. Although this was originally proposed to be used only on the discriminator, recent works \cite{zhang2018self, odena2018generator} suggest that the generator can also benefit from SN by suppressing sudden changes of parameter and gradient values. We apply SN to both the generator and discriminator. The final joint loss objective for $G_1$ with regularization parameters $\lambda_{G_1}$ and $\lambda_{FM}$ thus becomes
\begin{equation}
	\mathcal{J}_{G_1} = \lambda_{G_1} \mathcal{L}_{G_1} + \lambda_{FM} \mathcal{L}_{FM} ,
	\label{eq:g1}
\end{equation}
where we choose $\lambda_{G_1}$ = 1 and $\lambda_{FM} = 10$ for all experiments. 

\subsection{Image Completion}
The image completion stage upscales the LR image to an incomplete HR image as input to $G_2$ using a fixed fractionally strided convolution kernel. This has the effect of adding empty rows and columns in-between pixels. To offset the pixels and increase the size of an image by a factor of $s$ we use an $s \times s$ convolution kernel with stride of $1/s$. Let $K$ denote a fixed strided convolution kernel and $\mathbf{\hat{I}}^{HR}$ represent the high-resolution image being constructed by offsetting the pixels from the LR image. 

\begin{figure}[h]
	\begin{equation*}
	\begin{split}
	K_2 =
	\setlength\arraycolsep{4pt}
	\begin{bmatrix}
	1 & 0 \\
	0 & 0
	\end{bmatrix}
	\hspace{10mm}
	K_4 =
	\setlength\arraycolsep{3pt}
	\renewcommand{\arraystretch}{0.6}
	\begin{bmatrix}
	1 & 0 & 0 & 0 \\
	0 & 0 & 0 & 0 \\
	0 & 0 & 0 & 0 \\
	0 & 0 & 0 & 0 
	\end{bmatrix}
	\end{split}
	\end{equation*}
	\caption{Fixed fractionally strided convolution kernels to offset the pixels of the LR image and create an incomplete HR image for $\times 2$ and $\times 4$ SISR factors.}
	\label{fig:sr_kernels}
\end{figure}

\begin{equation} 
    \mathbf{\hat{I}}^{HR} = \mathbf{I}^{LR} * K.
\end{equation}
The HR image is then generated using $G_2$:
\begin{equation}
    \mathbf{I}_{(pred)} = G_2(\mathbf{\hat{I}}^{HR}, \mathbf{C}_{(pred)}).
\end{equation}
We proceed to train $G_2$ with another joint loss consisting of an $l_1$ loss, hinge loss, perceptual loss, and style loss. The hinge variant of the adversarial loss follows equations \ref{eq:g1loss} and \ref{eq:d1loss}
\begin{equation}
	\mathcal{L}_{G_2} = - \mathbb{E}_{\mathbf{C}_{pred}} \left[ D_2 (\mathbf{I}_{pred}, \mathbf{C}_{pred}) \right],
\end{equation}
\begin{multline}
	\mathcal{L}_{D_2} = \mathbb{E}_{(\mathbf{I}_{gt},\mathbf{C}_{pred})} \left[ \max (0, 1 - D_2 (\mathbf{I}_{gt}, \mathbf{C}_{pred})) \right] \\
	+ \mathbb{E}_{\mathbf{C}_{pred}} \left[ \max (0, 1 + D_2 (\mathbf{I}_{pred}, \mathbf{C}_{pred})) \right].
\end{multline} 
We include style loss $\mathcal{L}_{style}$ and perceptual loss $\mathcal{L}_{perc}$ \cite{gatys2016image, johnson2016perceptual} in our joint loss objective to further supplement training. Perceptual loss minimizes the Manhattan distance between feature maps generated from intermediate layers of VGG-19 trained on the ImageNet dataset \cite{russakovsky2015imagenet}. This has the effect of encouraging perceptually similar predictions with ground truth labels. Perceptual loss is defined as 

\begin{equation}
	\mathcal{L}_{perc} = \mathbb{E} \left[ \sum_{i} \frac{1}{N_i} \left\lVert \phi_i (\mathbf{I}_{gt}) - \phi_i (\mathbf{I}_{pred}) \right \rVert_1 \right],
\end{equation}
where $N_i$ is the number of elements in the $i$'th activation of VGG-19. While perceptual loss encourages perceptual similarities between ground truth images and predictions, style loss encourages texture similarities by minimizing the Manhattan distance between the Gram matrices of the intermediate feature maps. The Gram matrix of feature map $\phi_{i}$ is represented by $G_{j}^{\phi}$ \cite{gatys2016image} and distributes spatial information of texture, shape, and style. Style loss is defined as 

\begin{equation}
    \mathcal{L}_{style} = \mathbb{E} \left[ \sum_{j} \lVert G_j^{\phi} (\mathbf{I}_{gt}) - G_j^{\phi} (\mathbf{I}_{pred}) \rVert_1 \right].
\end{equation}
Style loss was shown by Sajjadi \etal \cite{sajjadi2017enhancenet} to successfully mitigate the ``checkerboard'' artifact caused by transpose convolutions \cite{odena2016deconvolution}. For both style and perceptual loss we extract feature maps from $\tt{relu1\_1}$, $\tt{relu2\_1}$, $\tt{relu3\_1}$, $\tt{relu4\_1}$ and $\tt{relu5\_1}$ of VGG-19. We do not use feature matching loss in the image completion stage. While the feature matching loss is a regularizer to the adversarial loss in the edge generator, the perceptual loss used in this stage has the same effect while it is shown to be more effective loss for image generation tasks \cite{nazeri2019edgeconnect,sajjadi2017enhancenet,johnson2016perceptual,johnson2016perceptual}. Thus the complete joint loss objective is  
\begin{equation}
	\mathcal{J}_{G_2} = \lambda_{\ell_1} \mathcal{L}_{\ell_1} + \lambda_{G_2} \mathcal{L}_{G_2} + \lambda_p \mathcal{L}_{perc} + \lambda_s \mathcal{L}_{style}.
	\label{eq:g2}
\end{equation}
In all of our experiments we choose to train with parameters $\lambda_{\ell_1} = 1$, $\lambda_{G_2} = \lambda_p = 0.1$, and $\lambda_s = 250$ to effectively minimize the reconstruction, style, perceptual, and adversarial loss to generate a photo-realistic high-resolution image.

\section{Experiments}
\subsection{Training Setup}
To train $G_1$, we generate edge maps using Canny edge detector \cite{canny1986computational}. We can control the level of detail in the LR edge map by changing the Gaussian filter smoothing parameter $\sigma$. For our purposes, we found $\sigma \approx 2$ yields the best results. All of our experiments are implemented in PyTorch, with the HR images fixed at $512\times512$ and the LR input scaled accordingly based on the zooming factor. We choose a batch size of eight during training. The models of both stages were optimized using Adam optimizer \cite{kingma2014adam} with $\beta_1 = 0$ and $\beta_2 = 0.9$. In our experiments, we didn't find any improvement by jointly optimizing $G_1$ and $G_2$, also we are limited to a smaller batch size due to the large memory footprint of the joint optimization, hence the generators from each stage are trained separately. We train $G_1$ using a learning rate of $10^{-4}$ with Canny edges until the loss plateaus. We lower the learning rate to $10^{-5}$ and continue training until convergence. We then freeze the weights of $G_1$ and continue to train $G_2$ with the same learning rates.

\subsection{Datasets}
Our proposed models are evaluated on the following publicly available datasets. 

\begin{itemize}
	\item Celeb-HQ \cite{karras2018progressive}. High-quality version of the CelebA dataset with 30K images. \\ \url{https://github.com/tkarras/progressive_growing_of_gans}
	
	\item Places2 \cite{zhou2017places}. More than 10 million images comprising 400+ unique scene categories. \\ \url{http://places2.csail.mit.edu/}
	
	\item Set5, Set14, BSDS100, Urban100 \cite{huang2015single}. Standard SISR evaluation datasets. \\ \url{http://vllab.ucmerced.edu/wlai24/LapSRN/}
\end{itemize}
\noindent
Results are compared against the current state-of-the-art methods both qualitatively and quantitatively.

\subsection{Qualitative Evaluation}
\begin{figure*}
	\centering
	\begin{subfigure}[c]{\textwidth}
		\centering
		\includegraphics[width=\textwidth]{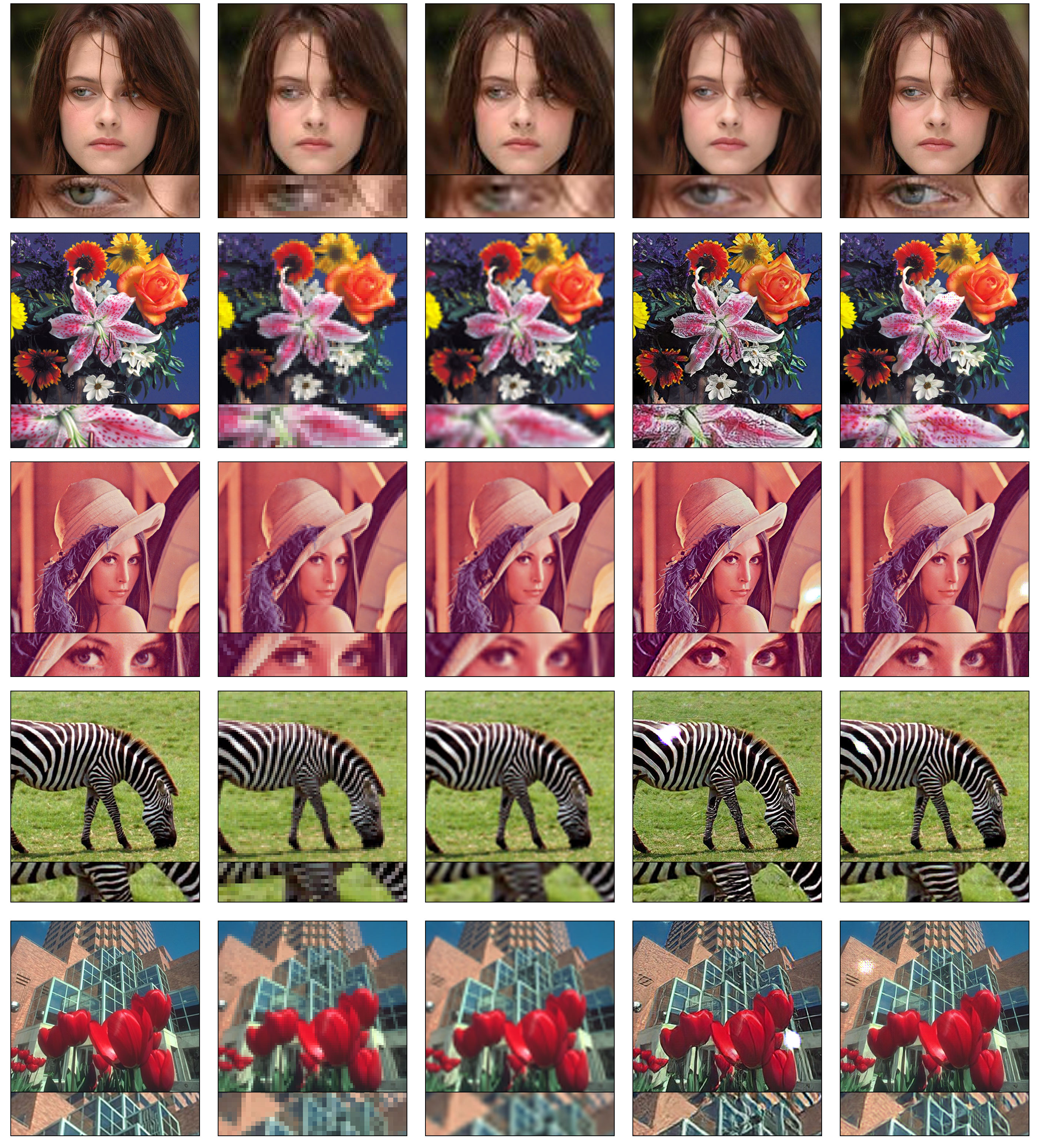}
		\caption*{\footnotesize \hspace{-7mm} Ground Truth \hspace{24mm} LR \hspace{28mm} Bicubic \hspace{66px} Baseline \hspace{71px} Ours}
	\end{subfigure}
	\caption{Comparison of qualitative results of images for $\times 4$ scale factor SISR cropped at $512 \times 512$. Left to right: Ground Truth HR, LR image upscaled using nearest-neighbor interpolation, SISR using bicubic interpolation, Baseline (no edge data), Ours (Full Model)}
	\label{fig:sisr_results_2x}
\end{figure*}

\begin{figure*}
	\centering
	\begin{subfigure}[c]{\textwidth}
		\centering
		\includegraphics[width=\textwidth]{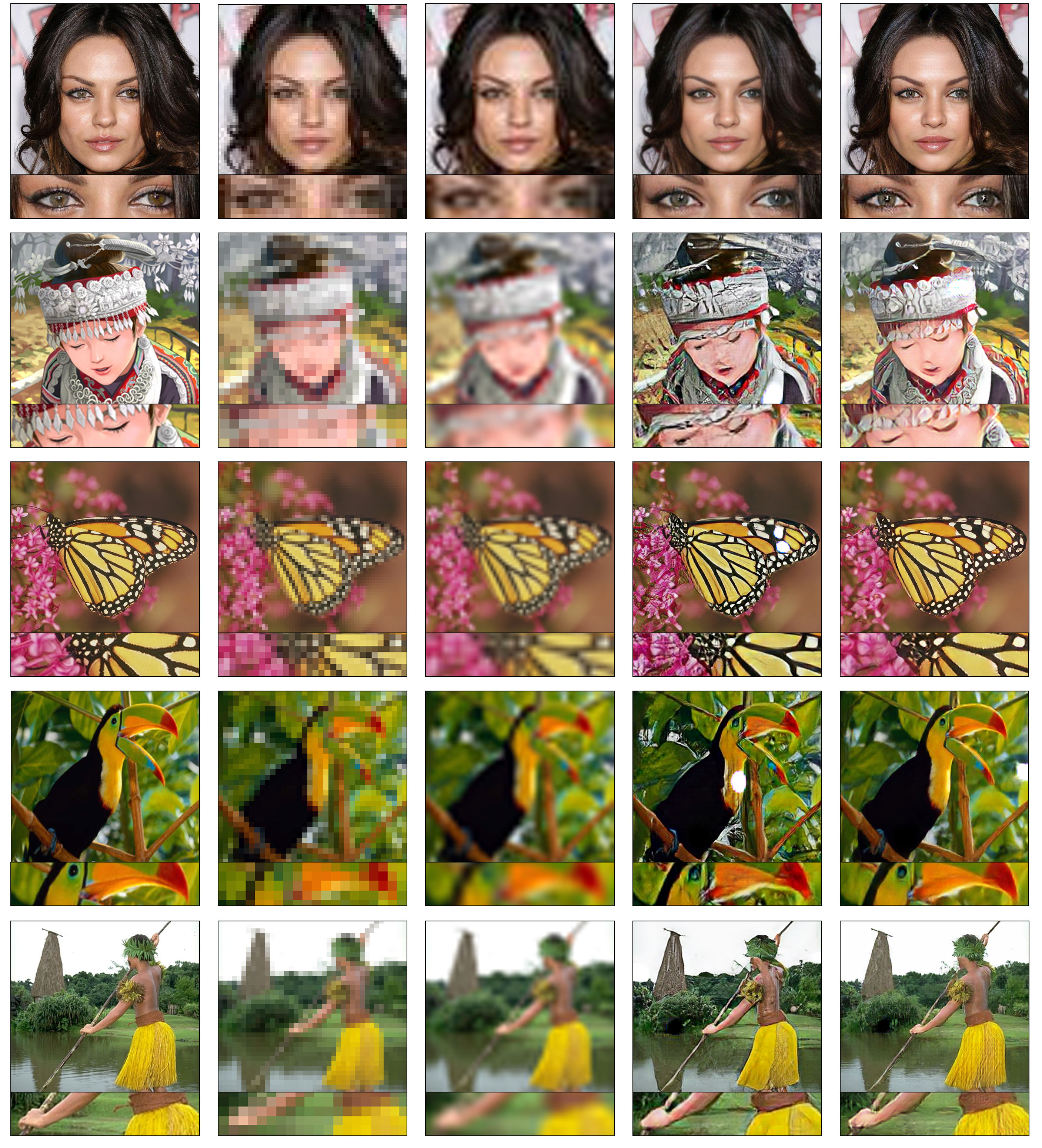}
		\caption*{\footnotesize \hspace{-7mm} Ground Truth \hspace{24mm} LR \hspace{28mm} Bicubic \hspace{66px} Baseline \hspace{71px} Ours}
	\end{subfigure}
	\caption{Comparison of qualitative results of images for $\times 8$ scale factor SISR cropped at $512 \times 512$. Left to right: Ground Truth HR, LR image upscaled using nearest-neighbor interpolation, SISR using bicubic interpolation, Baseline (no edge data), Ours (Full Model)}
	\label{fig:sisr_results_4x}
\end{figure*}

Figures \ref{fig:sisr_results_2x} and \ref{fig:sisr_results_4x} show results of the proposed SISR method for scale factors of $\times 4$ and $\times 8$ respectively. For visualization purposes, the LR image is resized using nearest-neighbor interpolation. All HR images are cropped at $512 \times 512$, which means the LR images are $128 \times 128$ and $64 \times 64$ for scale factors of $\times 4$ and $\times 8$ respectively. We obtain the LR images by blurring the HR with a Gaussian kernel of width $\sigma = 1$ followed by downsampling with the corresponding zooming scale factor. The results are compared against bicubic interpolation and our proposed model without the edge generation network as a baseline. Despite having almost high PSNR/SSIM, the baseline model produces blurry results around the edges while our full model (with edge-maps) remains faithful to the high-frequency edge data and produces sharp photorealistic images.

\begin{table*}
	\def\arraystretch{1.4}
	\centering
	\begin{tabular}{c|c*{6}{|>{\centering\arraybackslash}p{.12\linewidth}}}
		\multicolumn{3}{r|}{\textbf{Dataset}} & \small{Bicubic} & \small{ENet} & \small{EDSR} & \small{Baseline} & \small{Ours} \\ \hhline{*{7}{=|}=}
		\multirow{12}{*}{\rotatebox{90}{PSNR}}
		& \multirow{4}{*}{$\bm{\times 2}$}
		& \small{Set5} & 33.66 & 33.89 & \textbf{38.20} & 27.32 & 33.60 \\ \cline{3-8}
		& & \small{Set14} & 30.24 & 30.45 & \textbf{34.02} & 24.86 & 29.24 \\ \cline{3-8}
		& & \small{BSD100} & 29.56 & 28.30 & \textbf{32.37} & 23.97 & 28.12 \\ \cline{3-8}
		& & \small{Celeb-HQ} & \textbf{33.25} & - & - & 31.33 & 32.12 \\ \hhline{~|*{6}{=|}=}
		& \multirow{4}{*}{$\bm{\times 4}$}
		& \small{Set5} & 28.42 & 28.56 & \textbf{32.62} & 24.22 & 28.59 \\ \cline{3-8}
		& & \small{Set14} & 25.99 & 25.77 & \textbf{28.94} & 21.56 & 25.19 \\ \cline{3-8}
		& & \small{BSD100} & 25.96 & 24.93 & \textbf{27.79} & 20.78 & 24.25 \\ \cline{3-8}
		& & \small{Celeb-HQ} & \textbf{29.59} & - & - & 27.94 & 28.23 \\ \hhline{~|*{6}{=|}=}
		& \multirow{4}{*}{$\bm{\times 8}$}
		& \small{Set5} & \textbf{23.80} & - & - & 19.32 & 23.73 \\ \cline{3-8}
		& & \small{Set14} & \textbf{22.37} & - & - & 18.47 & 21.44 \\ \cline{3-8}
		& & \small{BSD100} & \textbf{22.11} & - & - & 18.65 & 21.63 \\ \cline{3-8}
		& & \small{Celeb-HQ} & \textbf{26.66} & - & - & 25.46 & 25.56 \\ \hhline{*{7}{=|}=}
		\multirow{12}{*}{\rotatebox{90}{SSIM}}
		& \multirow{4}{*}{$\bm{\times 2}$}
		& \small{Set5} & 0.930 & 0.928 & 0.961 & 0.974 & \textbf{0.985} \\ \cline{3-8}
		& & \small{Set14} & 0.869 & 0.862 & 0.920 & 0.930 & \textbf{0.954} \\ \cline{3-8}
		& & \small{BSD100} & 0.843 & 0.873 & 0.902 & 0.909 & \textbf{0.932} \\ \cline{3-8}
		& & \small{Celeb-HQ} & 0.967 & - & - & 0.957 & \textbf{0.968} \\ \hhline{~|*{6}{=|}=}
		& \multirow{4}{*}{$\bm{\times 4}$}
		& \small{Set5} & 0.810 & 0.809 & 0.898 & 0.929 & \textbf{0.965} \\ \cline{3-8}
		& & \small{Set14} & 0.703 & 0.678 & 0.790 & 0.832 & \textbf{0.894} \\ \cline{3-8}
		& & \small{BSD100} & 0.668 & 0.627 & 0.744 & 0.773 & \textbf{0.851} \\ \cline{3-8}
		& & \small{Celeb-HQ} & 0.834 & - & - & 0.910 & \textbf{0.912} \\ \hhline{~|*{6}{=|}=}
		& \multirow{4}{*}{$\bm{\times 8}$}
		& \small{Set5} & 0.646 & - & - & 0.801 & \textbf{0.904} \\ \cline{3-8}
		& & \small{Set14} &  0.552 & - & - & 0.708 & \textbf{0.793} \\ \cline{3-8}
		& & \small{BSD100} & 0.532 & - & - & 0.663 & \textbf{0752} \\ \cline{3-8}
		& & \small{Celeb-HQ} & 0.782 & - & - & 0.841 & \textbf{0.857} \\ \hline
	\end{tabular}
	\caption{Comparison of PSNR and SSIM for $\times 2$, $\times 4$, and $\times 8$ factor SISR over \textbf{Set5}, \textbf{Set14}, \textbf{BSD100}, and \textbf{Celeb-HQ} datasets with bicubic interpolation, ENet \cite{sajjadi2017enhancenet}, EDSR \cite{lim2017enhanced}, and baseline (without edge-data). The best result of each row is boldfaced.}
	\label{tab:sisr_num}
\end{table*}

\begin{figure*}
	\centering
	\begin{subfigure}[c]{\textwidth}
		\centering
		\includegraphics[width=\textwidth]{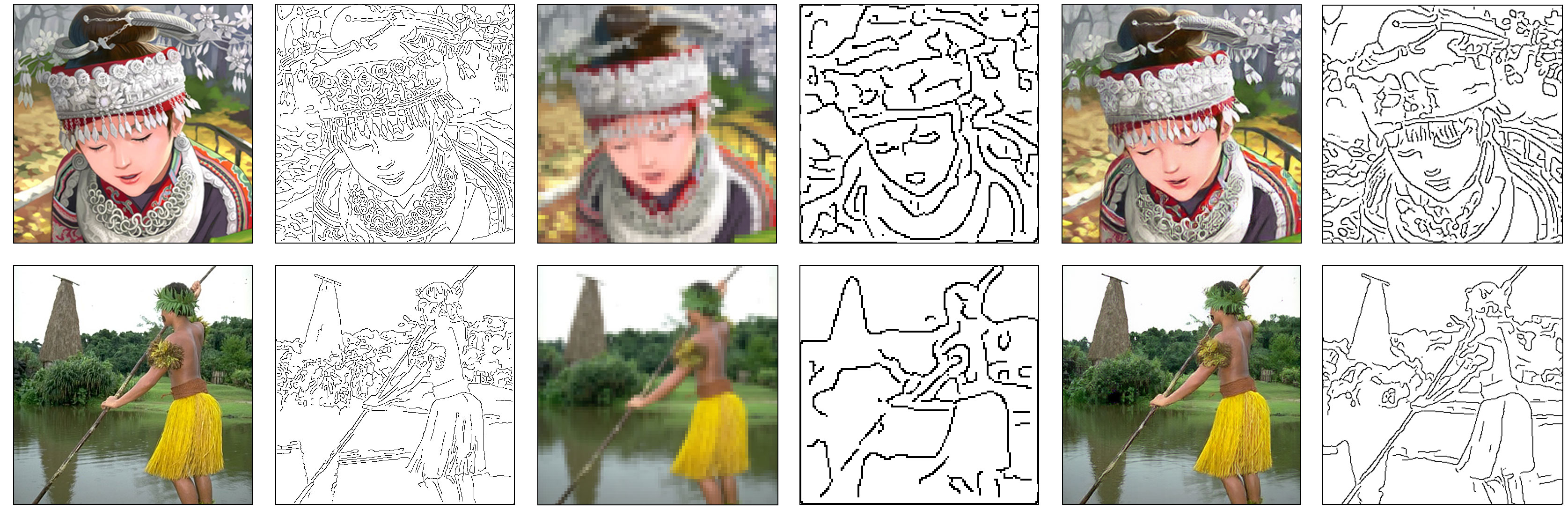}
		\caption*{\footnotesize \hspace{-5mm} Ground Truth \hspace{50mm} LR \hspace{145px} $\times 4$ SISR}
	\end{subfigure}
	\caption{Comparison of edge prediction results for $\times 4$ scale factor SISR cropped at $512 \times 512$. Left to right: Ground Truth HR, HR edge-map, LR image upscaled using nearest-neighbor interpolation, LR edge-map upscaled using nearest-neighbor interpolation, $\times 4$ SISR, $\times 4$ predicted edge-map SISR. }
	\label{fig:sisr_results_4x_edge}
\end{figure*}

\begin{table}
	\centering
	\def\arraystretch{1.2}
	\begin{tabular}{c*{3}{|>{\centering\arraybackslash}p{.2\linewidth}}}
		\multicolumn{2}{c|}{\textbf{Scale}} & Precision & Recall \\ \hhline{=*{3}{|=}}
		\multirow{3}{*}{\rotatebox{90}{Celeb-HQ}}
		& \bm{$\times 2$} & 74.27 & 73.21 \\ \cline{2-4}
		& \bm{$\times 4$} & 45.14 & 43.04 \\ \cline{2-4}
		& \bm{$\times 8$} & 23.23 & 19.09 \\ \hhline{=*{3}{|=}}
		\multirow{3}{*}{\rotatebox{90}{Places2}}
		& \bm{$\times 2$} & 79.18 & 80.24 \\ \cline{2-4}
		& \bm{$\times 4$} & 60.80 & 58.19 \\ \cline{2-4}
		& \bm{$\times 8$} & 31.06 & 23.93 \\ \hline
	\end{tabular}
	\caption{Quantitative performance of edge enhancer for Single Image Super-Resolution trained on Canny edges with $\sigma = 2$ for $512 \times 512$ images. Statistics are calculated over the standard test sets of each dataset.\\\\}
	\label{tab:sisr_acc}
\end{table}

\subsection{Quantitative Evaluation}
We evaluate our model using PSNR and SSIM for $\times 2$, $\times 4$ and $\times 8$ SISR scale factors. Table \ref{tab:sisr_num} shows the performance of our model against bicubic interpolation and current state of the art SISR models over datasets Set5, Set14, BSD100, and Celeb-HQ. Statistics for competing models for $\times 2$ and $\times 4$ SR were obtained from their respective papers where available. Results for a challenging case of $\times 8$ are only compared against bicubic interpolation. Note that the PSNR in our results is lower than competing models. In particular, EDSR by Lim \etal \cite{lim2017enhanced} has achieved the best PSNR for every dataset. However, their model is only trained with per-pixel $\ell_1$ loss and fails to reconstruct sharp edges despite having higher PSNR. Similar results in recent research \cite{johnson2016perceptual,sajjadi2017enhancenet} show that PSNR favors smooth/blurry results.

\subsection{Accuracy of Edge Generator}
Table \ref{tab:sisr_acc} shows the accuracy of our edge enhancer $G_1$ for Celeb-HQ and Places2 datasets for the Single Image Super-Resolution task. We measure precision and recall for various scale factors of SISR. In all experiments, the width of the Gaussian smoothing filter $\sigma = 2$ for Canny edge detection. 
\newline
Figure \ref{fig:sisr_results_4x_edge} shows results of the edge prediction stage for $\times 4$ scale factor. HR images are cropped at $512 \times 512$ and for visualization purposes, the LR image and its edge-map are resized using nearest-neighbor interpolation.

\section{Discussion and Future Work}
We propose a new structure-driven deep learning model for Single Image Super-Resolution (SISR) by recasting the problem as an in-between pixels inpainting task. One benefit of this approach over most deep-learning based SISR models is that we only have a unified model that can be used for different SISR zooming scales. Most deep-learning based SISR models take the LR image as input and generate the HR by in-network upsampling layers, given a zooming factor. For each different zooming factor, different network architecture and training is required. On the other hand, our model takes the LR image and adds empty space between pixels before using it as input to the network. Our proposed model learns to fill in the missing pixels by relying on available edge information to create the high-resolution image and effectively applies parameter sharing for different scales of SISR. Quantitative results show the effectiveness of the structure-guided inpainting model for the SISR problem where it achieves state-of-the-art results on standard benchmarks. 

One shortcoming of the proposed inpainting-based SISR model is that it requires minimizing two disjoint optimizing algorithms. A better approach is to incorporate the edge generation stage into the inpainting model's objective. This model could be trained using a joint optimization of image contents and structures and potentially outperform the disjoint two-stage optimization algorithm computationally while preserving sharp details of the image. 

Our method leads to an interesting direction, which raises the question that what other information could be learned from the original dataset to help the super-resolution process. Our source code is available at:\\ {\footnotesize\url{https://github.com/knazeri/edge-informed-sisr}}
%We leave these limitations as an interesting direction for future works.
%Another limitation of this model is that it is not easily scalable to non-integer scale factors. One solution would be to use an %interpolation method on a low-resolution image as well as the inpainting mask as inputs to the network. 
%In this scheme, the inpainting mask is no longer a binary mask but a heat-map that guides the completion network through the inpainting %process. 
%-------------------------------------------------------------------------
\section*{Acknowledgments} This research was supported in part by the Natural Sciences and Engineering Research Council of Canada (NSERC). We gratefully acknowledge the support of NVIDIA Corporation with the donation of the Titan V GPU used for this research.
\newpage
\bibliographystyle{ieee}
\bibliography{refs}

\end{document}